\documentclass[pra,twocolumn,aps,showpacs,floatfix,superscriptaddress]{revtex4-1}

\usepackage{graphicx}
\usepackage{rotating}
\usepackage{amsmath}
\usepackage{bbm}
\usepackage{subfigure}
\usepackage{color}
\usepackage{tabularx}

\newcommand{\be}{\begin{equation}}
\newcommand{\ee}{\end{equation}}
\newcommand{\bea}{\begin{eqnarray}}
\newcommand{\eea}{\end{eqnarray}}
\newcommand{{\br}}{\bf r}

\newcolumntype{L}[1]{>{\hsize=#1\hsize\raggedright\arraybackslash}X}%
\newcolumntype{R}[1]{>{\hsize=#1\hsize\raggedleft\arraybackslash}X}%
\newcolumntype{C}[2]{>{\hsize=#1\hsize\columncolor{#2}\centering\arraybackslash}X}%

\graphicspath{
  {./}
  {figures/pdf/}
  {figures/eps/}
  {figures/jpg/}
}

\begin{document}

\title{Generalized Pauli constraints in reduced density matrix functional theory}
\author{Iris Theophilou}
\affiliation{Peter-Gr\"unberg Institut and Institute for Advanced Simulation,
Forschungszentrum J\"ulich, D-52425 J\"ulich, Germany}
\author{Nektarios N.\ Lathiotakis}
\affiliation{Theoretical and Physical Chemistry Institute, National Hellenic 
Research Foundation, Vass.\  Constantinou 48, GR-11635 Athens, Greece}
\affiliation{Max-Planck-Institut f\"ur Mikrostrukturphysik, Weinberg 2, D-06120 Halle (Saale), Germany}
\author{Miguel A.L.\ Marques}
\affiliation{Institut f\"ur Physik Martin-Luther-Universit\"at Halle-Wittenberg, D-06120 Halle (Saale), 
Germany}
\author{Nicole Helbig}
\affiliation{Peter-Gr\"unberg Institut and Institute for Advanced Simulation,
Forschungszentrum J\"ulich, D-52425 J\"ulich, Germany}

\begin{abstract}  
\noindent
Functionals of the one-body reduced density matrix (1-RDM) are routinely minimized under Coleman's
ensemble $N$-representability conditions. Recently, the topic of pure-state $N$-representability conditions, 
also known as generalized Pauli constraints, received increased attention following the discovery of a systematic way to 
derive them for any number of electrons and any finite dimensionality of the Hilbert space. The target of this work 
is to assess the potential impact of the enforcement of the pure-state conditions on the results of reduced density-matrix functional theory calculations.  In particular, we examine whether the standard minimization of typical 1-RDM functionals under the ensemble $N$-representability conditions violates the pure-state conditions for prototype 3-electron systems. We also enforce the pure-state conditions, in addition to the ensemble ones, for the same systems and functionals and compare the correlation energies and optimal occupation numbers with those obtained by the enforcement of the ensemble conditions alone.
\end{abstract}

\date{\today}

\maketitle


\section{Introduction\label{sec:intro}}

In Reduced Density Matrix Functional Theory \cite{G1975} (RDMFT), the total energy of an 
$N$-electron system is expressed as a functional of the one-body reduced density-matrix (1-RDM) $\gamma(\br,\br')$
or its eigenfunctions $\phi_{j\sigma}(\br)$ and eigenvalues $n_{j\sigma}$, i.e.\ the 
natural orbitals and their occupation numbers. To obtain the 1-RDM of the ground state, the total 
energy is typically minimized under Coleman's ensemble $N$-representability conditions (EC) \cite{Coleman_1963},
\be\label{eq:ensnrep}
\sum_{j=1}^\infty n_{j\sigma}=N_\sigma, \quad 0\leq n_{j\sigma}\leq 1\,.
\ee
These conditions ensure that the 1-RDM corresponds to an ensemble of fermionic states. 
The advantage of the EC is their simplicity
when compared with the pure-state $N$-representability conditions (PC). The PC are the necessary and 
sufficient conditions for the 1-RDM to correspond to a pure state and are also known as generalized Pauli 
conditions\cite{Klyachko_Math,Klyachko_JPhys}.

Although some of the PC were known for a long time\cite{Borland-Dennis,Smith}, 
a systematic way to derive all conditions for a given number of particles $N$ and finite size
$M$ of the Hilbert space (being equal to the number of natural spin-orbitals) 
has been recently demonstrated by Klyachko et al. \cite{Klyachko_Math,Klyachko_JPhys}. 
Since the number of natural spin-orbitals needs to be finite in any implementation, these 
conditions can be derived for all practical purposes. As the conditions depend on 
both $N$ and $M$, we will refer to them as the $N$-$M$ conditions in the following. In practice, the number of PC 
increases rapidly with the number 
of electrons and natural spin-orbitals. For example, there are four 3-6 conditions while the number 
of 3-8 conditions is already 31. All the PC can be written as inequalities for various subsets of the occupation numbers. Geometrically, the set of conditions defines a $M$-dimensional convex polytope. Investigations into the relevance of the conditions for physical systems have started recently\cite{PRL_pure_cond,Benavides,Mazziotti_pure,Schilling_2015} focussing on the question whether the solutions lie on the boundary of the $M$-dimensional polytope or inside. In this work, we discuss the PC in the context of approximate functionals in RDMFT calculations.

Several approximations of the total energy as an explicit functional of the natural orbitals and the
occupation numbers (or the 1-RDM) have been introduced in the last 
decades \cite{M1984,BB2002,GU1998,GPB2005,RPGB2008,ML2008,LM2008,SDLG2008,LSDEMG2009,P2006,PMLU2010,PLRMU2011,P2014,GGB2012,S1999,R2002,RX2007,KP2014}
and employed in electronic
structure calculations in combination with the EC. In these approximations, the exact kinetic and external
 potential energy functionals in terms of the 
1-RDM are employed while the electron-electron interaction is approximated. Among these 
approximations a central position is held by the M\"uller functional \cite{M1984,BB2002}
 where $\gamma$ in the exchange term of Hartree-Fock theory is replaced by $\gamma^{1/2}$ 
(power in the operator sense). This approximation was shown to 
overestimate correlation energies substantially and several
remedies were put forward. For example, Goedecker and Umrigar \cite{GU1998}, proposed a
modified form of the M\"uller functional where self-interaction terms (same index terms) 
are explicitly excluded from the double index summations of the Coulomb and exchange like energy
terms.  

Gritsenko et al.\ \cite{GPB2005} introduced a hierarchy of repulsive corrections
based on the separation of natural orbitals into strongly and weakly occupied
as well as bonding and antibonding orbitals. The proposed BBC$n$ ($n$ = 1,2,3) functionals, and in particular BBC3, 
reproduce quite accurately correlation energies and the 
dissociation of diatomic molecules. A modelling of the BBC3 approximation using a fitting formula 
was introduced by Rohr et al.\ \cite{RPGB2008} leading to the functional known as
Automatic Correction 3 (AC3).  

Marques and Lathiotakis \cite{ML2008} (ML) introduced an empirical functional by replacing the square root 
dependence on the occupations of the M\"uller functional with a Pad\'e-type formula with parameters fitted 
to reproduce the correlation energies of a test set of molecules. Another approximation, called the ``Power functional'', 
targeting the study of periodic systems, was proposed by Sharma et al.\ \cite{SDLG2008}, by substituting the square-root 
dependence on the occupations of the M\"uller functional with a power $\alpha$. A value of $\alpha= 0.578$
was found to optimize the performance for the functional for both finite and extended systems \cite{LSDEMG2009}.
 This functional was shown to reproduce quite successfully the band gaps of semiconductors and 
insulators including transition metal oxides \cite{SDLG2008,SDSG2013}. 

Based on the reconstruction of the two-body reduced density matrix in terms of the 1-RDM, Piris and coworkers developed a series of approximations \cite{P2006,PMLU2010,PLRMU2011,P2014} termed  PNOF$n$, $n$ = 1,...6 which were proven to be  accurate in reproducing several properties of molecular systems. Functionals that one could classify as extensions of RDMFT are those that depend explicitly on the phases of natural orbitals \cite{GGB2012} like for example the exact functional for 2-electron systems of L\"owdin and Shull \cite{SL2956} and the theory of the anti-symmetrized product of strongly orthogonal geminals (APSG) \cite{S1999}. APSG accurately predicts structural and vibrational properties of molecules as well as single-bond breaking \cite{R2002,RX2007,KP2014}. 

Recently, and in order to address the large numerical cost of the orbital minimization in RDMFT, local-RDMFT was introduced \cite{LHRG2014a,LHRG2014b} which combines the non-idempotent 1-RDMs of RDMFT with a local, Kohn-Sham-like potential in a variational manner. 

One could argue that, in the minimization of the total energy in RDMFT, PC should be enforced 
in addition to the EC. For the exact functional, however, due to the minimization principle, the EC
are sufficient to guarantee that the minimizing 1-RDM corresponds to a pure state, i.e.\ the ground state, in all cases 
without ground-state degeneracy. For approximate functionals, however, this is not the case. In other words, 
in the minimization of an approximate functional of the 1-RDM, if one applies the EC alone or the PC in addition, 
there is no guarantee that the optimal 1-RDMs will coincide.

In this paper, we assess how well different approximate RDMFT functionals, combined with only the EC,
satisfy pure-state conditions for different 3-electron systems, namely the lithium atom, the LiH$^+$ 
and the He$_2^+$ ions, and the hydrogen trimer, using 6 or 8 natural spin-orbitals in the 
calculation. In addition, we perform RDMFT calculations for these 3-electron systems enforcing 
the PC as additional constraints and compare the results with and without enforcing them.
The goal of this work is to get a first feeling on whether imposing the PC has any impact on the RDMFT results or if the 
EC are sufficient. Three electron systems are the smallest systems where one can investigate this question because for closed-shell systems with time-reversal symmetry the EC are proven to be necessary and sufficient for pure-state $N$-representability \cite{Smith}.

In section~\ref{sec:psnrep}, we briefly review the known conditions for pure-state $N$-representability for 
3-electron systems and for Hilbert spaces of 6 and 8 natural spin-orbitals. We discuss our results for the four different systems with and without enforcing the PC in section~\ref{sec:results} and conclude our 
findings in section~\ref{sec:conclusion}.

\section{Pure state $N$-representability conditions}\label{sec:psnrep}

Similarly to the EC, the conditions for the pure-state $N$-representability
concern only the occupation numbers. To express the PC, at least in the form known so far,
occupation number ordering according to their value is essential.
Thus, in order to write the PC we have to map the set of occupation numbers $\{n_{j\sigma}\}$
onto an ordered set $\{\lambda_k\}$, $\lambda_1\geq \lambda_2\geq...\geq\lambda_M$ regardless of the
spin they correspond to. As a realistic example considered in this work, for the lithium atom, the M\"uller 
functional with the EC alone and 6 natural spin-orbitals, we obtain
\be
\begin{array}{lll}
n_{1\uparrow}= 1.000000, & n_{2\uparrow}=0.865779, & n_{3\uparrow}=0.134220,\\
n_{1\downarrow}=0.999969, & n_{2\downarrow}=0.000027, & n_{3\downarrow}=0.000005,
\end{array}
\ee
which are then mapped according to
\be
\label{eq:mapli}
\begin{array}{lll}
n_{1\uparrow}\rightarrow\lambda_1, & n_{2\uparrow}\rightarrow\lambda_3, & n_{3\uparrow}\rightarrow\lambda_4\\
n_{1\downarrow}\rightarrow\lambda_2, & n_{2\downarrow}\rightarrow\lambda_5, & n_{3\downarrow}\rightarrow\lambda_6\,.
\end{array}
\ee
We note that using three occupation numbers for each spin channel is a question of choice. One could choose an unbalanced distribution of the 6 occupation numbers into the two channels.
For systems of 3 electrons, and for a Hilbert space of 6 natural spin-orbitals, there are 4 conditions which are 
given in Table \ref{tab:36const}. The first three of the conditions are originally inequalities, like all Klyachko's conditions, with the left-hand-side being less or equal to one, however, they reduce to equalities due to the ensemble $N$-representability condition Eq.\ (\ref{eq:ensnrep}) which the occupation numbers have to satisfy as well. As one can see, for the example above, only the first and the third conditions are well satisfied while the second is violated slightly and the forth quite severely.

The conditions in Table \ref{tab:36const} were originally formulated by Borland and Dennis in the 70's \cite{Borland-Dennis}. For a 3 electron wavefunction expressed in terms of 6 natural spin-orbitals, $\phi_1...\phi_6$, (three for each spin channel) one can form 9 Slater determinants that could contribute to the wavefunction. Borland and Dennis found numerically that only four of these determinants have non-zero coefficients. These determinants are $|\phi_1 \phi_2 \phi_3|$, $|\phi_1 \phi_4 \phi_5|$, $|\phi_2 \phi_4 \phi_6|$ and $|\phi_3 \phi_5 \phi_6|$. We label the  expansion coefficients of these determinants with $c_1$, $c_2$, $c_3$, and $c_4$, respectively. Forming the corresponding 1-RDM we can associate every occupation number $\lambda_i$ with two expansion coefficents $c_i$ because each orbital is present in only two Slater determinants. For example, $\lambda_1=|c_1|^2+|c_2|^2$ and $\lambda_6=|c_3|^2+|c_4|^2$. Since the wavefunction has to be normalized, $\lambda_1 +\lambda_6$ has to be one, which is the first Borland - Dennis condition. In the same way, one obtains the second and third conditions. Moreover, every $|c_i|^2$ has to be non-negative from which the forth condition follows. In this way, Borland and Dennis proved the sufficiency of their conditions and referred to a work by M.B.\ Ruskai and R.L.\ Kingsley for their necessity. This work, however, was only published 35 years later by M.B.\ Ruskai \cite{Ruskai}.

\begin{table}
\begin{tabular}{r|l}
\# & Condition\\ \hline
1 & $\lambda_1+\lambda_6 = 1$,\\
2 & $\lambda_2+\lambda_5 = 1$,\\
3 & $\lambda_3+\lambda_4 = 1$,\\
4 & $\lambda_5+\lambda_6-\lambda_4 \geq 0$.
\end{tabular}
\caption{\label{tab:36const} 3-6 PC, i.e.\ for 3 electrons in a space of 6 natural spin-orbitals \cite{Borland-Dennis}.}
\end{table}

The number of conditions increases dramatically with the number of natural spin-orbitals that one includes in the calculation. 
Using invariant theory and represenation theory, Klyachko and coworkers derived and proved an algorithm for computing all such Pauli-like constraints for a given number of electrons and natural spin-orbitals. For the case of 8 orbitals they obtained 31 conditions that need to be enforced which are shown in Table~\ref{tab:38const} \cite{ThesisAltunbulak}.

\begin{table}
\begin{tabular}{r|l}
\# & Condition\\ \hline
1  & $\lambda_1+\lambda_2+\lambda_4+\lambda_7\leq 2$\\
2  & $\lambda_1+\lambda_2+\lambda_5+\lambda_6\leq 2$\\ 
3  & $\lambda_2+\lambda_3+\lambda_4+\lambda_5\leq 2$\\
4  & $\lambda_1+\lambda_3+\lambda_4+\lambda_6\leq 2$\\ \hline
5  & $\lambda_1+\lambda_2-\lambda_3\leq 1$\\
6  & $\lambda_2+\lambda_5-\lambda_7\leq 1$\\
7  & $\lambda_1+\lambda_6-\lambda_7\leq 1$\\
8  & $\lambda_2+\lambda_4-\lambda_6\leq 1$\\
9  & $\lambda_1+\lambda_4-\lambda_5\leq 1$\\
10 & $\lambda_3+\lambda_4-\lambda_7\leq 1$\\ \hline
11 & $\lambda_1+\lambda_8\leq 1$\\ \hline
12 & $\lambda_2-\lambda_3-\lambda_6-\lambda_7\leq 0$\\
13 & $\lambda_4-\lambda_5-\lambda_6-\lambda_7\leq 0$\\
14 & $\lambda_1-\lambda_3-\lambda_5-\lambda_7\leq 0$\\ \hline
15 & $\lambda_2+\lambda_3+2\lambda_4-\lambda_5-\lambda_7+\lambda_8\leq 2$\\
16 & $\lambda_1+\lambda_3+2\lambda_4-\lambda_5-\lambda_6+\lambda_8\leq 2$\\
17 & $\lambda_1+2\lambda_2-\lambda_3+\lambda_4-\lambda_5+\lambda_8\leq 2$\\
18 & $\lambda_1+2\lambda_2-\lambda_3+\lambda_5-\lambda_6+\lambda_8\leq 2$\\ \hline
19 & $\lambda_1+\lambda_2-2\lambda_3-\lambda_4-\lambda_5\leq 0$\\
20 & $\lambda_1-\lambda_2-\lambda_3+\lambda_6-2\lambda_7\leq 0$\\ \hline
21 & $\lambda_1-\lambda_3-\lambda_4-\lambda_5+\lambda_8\leq 0$\\
22 & $\lambda_1-\lambda_2-\lambda_3-\lambda_7+\lambda_8\leq 0$\\ \hline
23 & $2\lambda_1-\lambda_2+\lambda_4-2\lambda_5-\lambda_6+\lambda_8\leq 1$\\
24 & $\lambda_3+2\lambda_4-2\lambda_5-\lambda_6-\lambda_7+\lambda_8\leq 1$\\
25 & $2\lambda_1-\lambda_2-\lambda_4+\lambda_6-2\lambda_7+\lambda_8\leq 1$\\
26 & $2\lambda_1+\lambda_2-2\lambda_3-\lambda_4-\lambda_6+\lambda_8\leq 1$\\
27 & $\lambda_1+2\lambda_2-2\lambda_3-\lambda_5-\lambda_6+\lambda_8\leq 1$\\ \hline
28 & $2\lambda_1-2\lambda_2-\lambda_3-\lambda_4+\lambda_6-3\lambda_7+\lambda_8\leq 0$\\
29 & $-\lambda_1+\lambda_3+2\lambda_4-3\lambda_5-2\lambda_6-\lambda_7+\lambda_8\leq 0$\\
30 & $2\lambda_1+\lambda_2-3\lambda_3-2\lambda_4-\lambda_5-\lambda_6+\lambda_8\leq 0$\\
31 & $\lambda_1+2\lambda_2-3\lambda_3-\lambda_4-2\lambda_5-\lambda_6+\lambda_8\leq 0$
\end{tabular}
\caption{\label{tab:38const} 3-8 PC, i.e.\  for 3 electrons in a space of 8 natural orbitals. 
Horizontal lines distinguish different types of conditions \cite{ThesisAltunbulak}.}
\end{table}

\section{Results}\label{sec:results}

\begin{figure}
\includegraphics[width=0.99\columnwidth, clip]{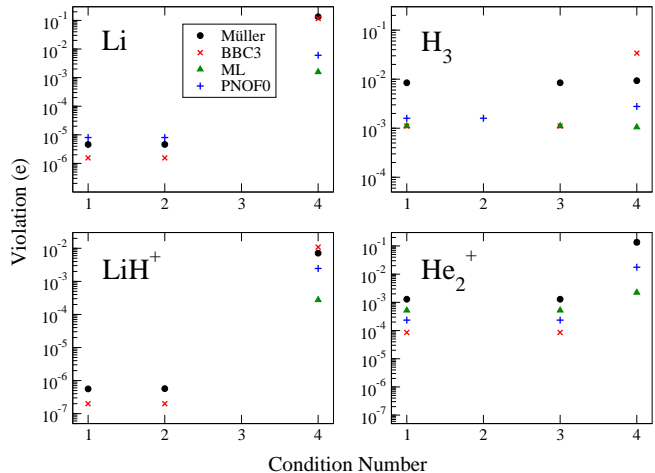}
\caption{\label{fig:36all} Violation in electrons (e) of the 3-6 PC for Li, H$_3$, LiH$^{+}$, and He$_{2}^{+}$,
with different RDMFT functionals minimized under the EC alone. On the horizontal axis, the conditions are numbered in the order shown in Table~\ref{tab:36const}.  A Missing symbol implies that the particular condition is not violated by the corresponding functional.}
\end{figure}

We perform all the calculations using the cc-PVTZ Gaussian basis set with the HIPPO computer code \cite{code}.
For the 3-electron doublet state we used the assumption that the natural spin-orbitals appear in pairs with identical spatial dependence \cite{LHG2005}, i.e.\ the spin dependence enters only through the occupation numbers. This restriction
is not expected to influence the findings of the present work since both PC and EC concern the occupation numbers only.  
We performed two kinds of calculations: (i) standard RDMFT calculations using the 
M\"uller \cite{M1984,BB2002}, BBC3 \cite{GPB2005}, ML \cite{ML2008} and PNOF0 \cite{P2006,LM2008}
functionals to examine
whether, and to what extend, PC are satisfied by the optimal occupations, and (ii) RDMFT calculations
with the 3-6 or 3-8 PC enforced in addition. 
We note that only 3 or 4 spatial orbitals (6, 8 natural spin-orbitals) were assumed to have non-zero
occupation even in case (i) in order to conform with the Hilbert space restrictions of the PC
and to allow for fair comparisons. We also note that the large size of the basis set allows for the use of the optimal $M$-dimensional Hilbert space in each calculation.
Occupation number and orbital optimizations were performed with customized sequential quadratic and conjugate gradient
methods, respectively. Our calculations are for testing purposes, since 6 spin-orbitals 
are generally not enough to describe dynamic correlations in 3-electron systems.
However, they are still reasonable because for 8 spin-orbitals the smallest occupation number
is typically of the order of 10$^{-5}$-10$^{-6}$.


We first examine whether the occupation numbers obtained with standard RDMFT calculations, i.e.\ by enforcing only the EC, satisfy the PC. In Fig.\ \ref{fig:36all}, we show the violation of the 3-6 PC in electrons for Li, LiH$^{+}$, He$_{2}^{+}$, and H$_{3}$. As the first 3 
of the 3-6 conditions are equalities the quantity we refer to as ``violation'' is the sum of the 
relevant occupation numbers minus 1 (the rhs.\ of all equalities). The fourth condition is violated when 
$\lambda_5 +\lambda_6 -\lambda_4<0$ and the violation is the difference of the lhs.\ of the condition from zero, i.e.\ the minimum value of the lhs.\ satisfying the condition. In a similar way, we determine the violation of the 3-8 conditions.
For the 3-6 conditions, we found that, for  all functionals and all systems apart from H$_{3}$, the fourth condition 
is always violated and has the highest violation. 
While for H$_3$ all violated conditions are of the same order of magnitude, for the rest of the systems
the fourth condition is violated the most by at least an order of magnitude. In all cases, the ML functional appears to 
give the smallest violations among the approximations employed.
Interestingly, when both the first and the second conditions are violated, as in Li and LiH$^+$, they are violated by the same amount. Due to the enforcement of the ensemble $N$-representability conditions and the pinning of the largest occupation number to one, the mapping (\ref{eq:mapli}) implies that both the first and the second conditions reduce to 
$n_{3\downarrow}=0$. Hence, they are violated by the same amount. For the hydrogen trimer and the He$_2^+$ molecule, the mapping onto ordered occupation numbers is different than in Eq.\ (\ref{eq:mapli}). As a result, the first and the third conditions reduce to $n_{3\downarrow}=0$ and are, hence, equally violated.
For Li and LiH$^+$, due to the pining of the first occupation number, $\lambda_1 = 1$, the third 
condition is automatically satisfied since the total number of electrons with spin up equals
$\lambda_1 + \lambda_3 + \lambda_4=2$.

\begin{figure*}
\begin{center}
\includegraphics[width=0.95\textwidth, clip]{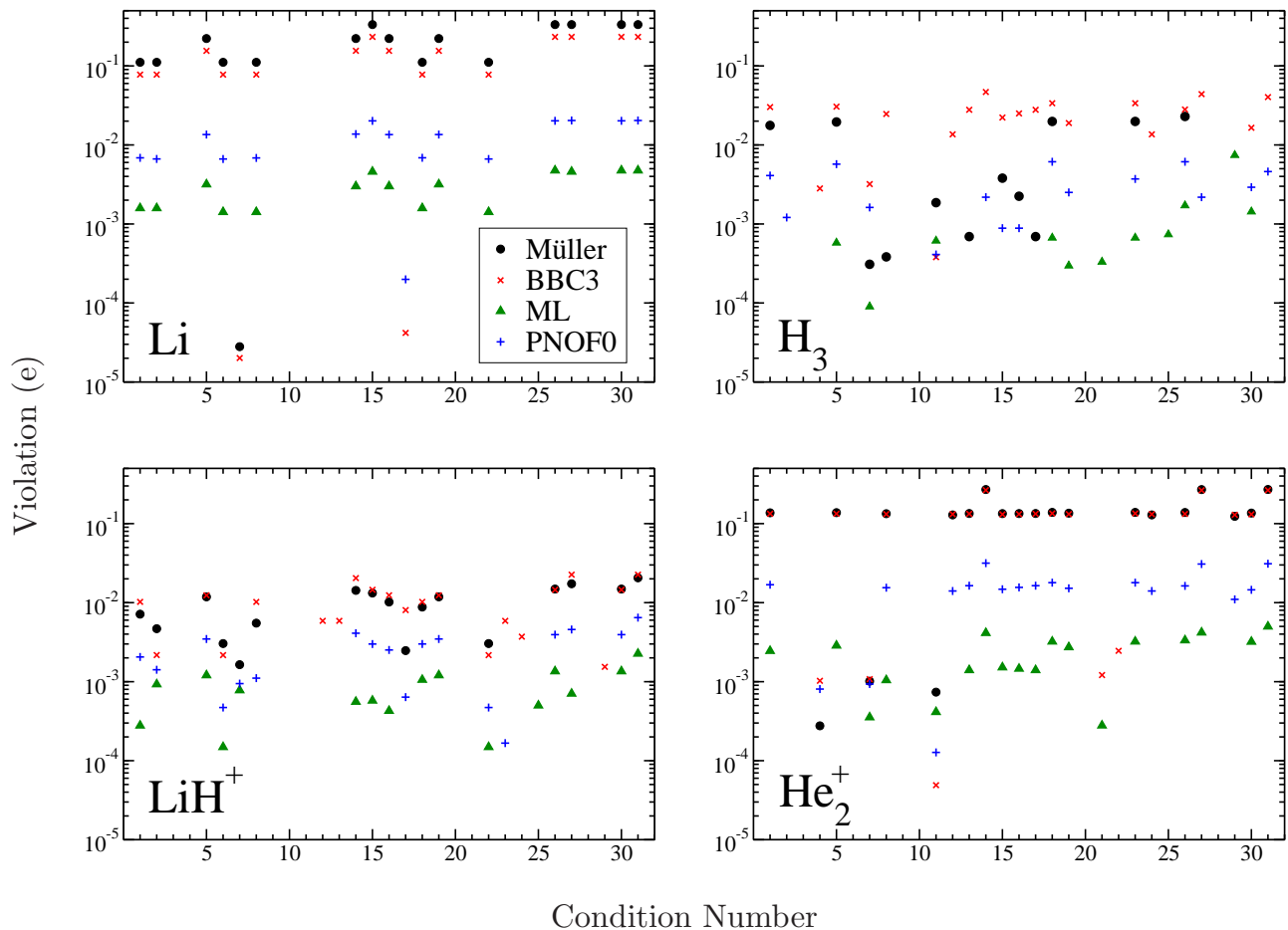}
\end{center}
\caption{\label{fig:38all} Violation in electrons (e) of the 3-8 PC for the Li atom, and the H$_3$, LiH$^{+}$, and He$_{2}^{+}$ molecules
with different RDMFT functionals minimized under the EC alone.
On the horizontal axis, the conditions are numbered in the order 
shown in Table~\ref{tab:38const}. A Missing symbol implies that the 
particular condition is not violated by the corresponding functional.}
\end{figure*}

In Fig.~\ref{fig:38all}, we show the violation of the 3-8 conditions. As we see, the conditions with numbers 9, 10, 20, and 28 in Table~\ref{tab:38const} are always satisfied while those with
numbers 5, 18, and 26 are always violated. For weakly correlated systems, where $\lambda_1$ to $\lambda_3$ are close to one and all remaining occupations are close to zero, it is clear that conditions 20 and 28 will always be satisfied. Condition 5 is always violated because we find that $\lambda_1$ is pinned to one for all systems and functionals considered here, which implies that the only way to satisfy the condition is $\lambda_2=\lambda_3$ which is actually not the case. For the  set of 3-8 conditions, again the ML functional gives the smallest violation, 
with PNOF0 coming next, while between the M\"uller and BBC3 functionals there is no clear order. These two functionals, however, 
show a similar violation for many of the conditions for all systems except H$_3$. Due to the larger number of conditions there are also several conditions that are violated to a similar extend, i.e.\ contrary to the 3-6 case there is not one dominant 
violation.

\begin{table}
\setlength{\tabcolsep}{2pt}
\begin{tabular}{l|cc|cc|cc|cc} \hline\hline
\multicolumn{9}{l}{\em A. For the 3-6 conditions:} \\ \hline
& \multicolumn{2}{c|}{M\"uller} & \multicolumn{2}{c|}{BBC3} & \multicolumn{2}{c|}{ML} &
\multicolumn{2}{c}{PNOF0} \\
& $|E_{c}^{\rm (E)}|$ & $\Delta E_{c}$ & $|E_{c}^{\rm (E)}|$ & $\Delta E_{c}$ & $|E_{c}^{\rm (E)}|$ & $\Delta E_{c}$ & $|E_{c}^{\rm (E)}|$ & $\Delta E_{c}$\\[1.2ex] \hline
Li      & 12.2 & 11.0 & 10.0 & 9.2 & 2.4 & 1.2 & 2.8 & 2.0\\
H$_3$ & 45.5 & 2.1 & 24.1 & 2.5 & 17.1 & 0.5 &20.2 & 0.5\\ 
LiH$^+$ & 16.1 & 1.5 & 7.7 & 5.6 & 19.1 &15.5  & 11.8 & 0.5 \\
He$_2^+$ & 81.7 & 39.1 & 71.4 & 43.3 & 25.8 & 2.0 & 32.8 & 7.5\\ \hline 
\multicolumn{9}{l}{}\\
\multicolumn{9}{l}{\em B. For the 3-8 conditions:} \\ \hline
& \multicolumn{2}{c|}{M\"uller} & \multicolumn{2}{c|}{BBC3} & \multicolumn{2}{c|}{ML} &
\multicolumn{2}{c}{PNOF0} \\
& $|E_{c}^{\rm (E)}|$ & $\Delta E_{c}$ & $|E_{c}^{\rm (E)}|$ & $\Delta E_{c}$ & $|E_{c}^{\rm (E)}|$ & $\Delta E_{c}$ & $|E_{c}^{\rm (E)}|$ & $\Delta E_{c}$\\[1.2ex] \hline
Li      & 23.4 & 20.0 & 16.5 & 15.2 &  4.8 & 1.3 & 5.6 & 3.6 \\
H$_3$   & 53.8 &  0.7 & 29.1  &  2.3 & 23.3 & 0.2 & 25.0 & 0.4 \\
LiH$^+$ & 19.9 &  1.0 & 10.3 &  6.2 & 22.2 & 1.8 & 15.2 & 3.2 \\
He$_2^+$& 90.7 & 38.0 & 76.6 & 42.4 & 34.3 & 1.3 & 37.9 & 6.3 \\
\hline\hline
\end{tabular}
\caption{\label{tab:corr} Absolute correlation energy, $|E_{c}^{\rm (E)}|$, for the EC
using 6 or 8 natural spin-orbitals, and its change, $\Delta E_{c}$, when the 3-6 or 3-8 PC are 
enforced in addition. All energies are given in mHa.}
\end{table}

In order to assess the impact of the PC on RDMFT calculations, we performed energy minimizations implementing the PC as 
additional constraints. A sensible comparison is that of the correlation energies $E_c^{\rm (E)}$ for the EC alone
and $E_c^{\rm (P)}$ when the PC are additionally enforced, defined as $E_c^{\rm (E)}=E^{\rm (E)}-E^{\rm (HF)}$ and 
$E_c^{\rm (P)}=E^{\rm (P)}-E^{\rm (HF)}$, where $E^{\rm (E)}$, $E^{\rm (P)}$ are the total energies obtained with the 
enforcement of the EC alone,
and the PC in addition, respectively, and $E^{\rm (HF)}$ is the Hartree-Fock total energy. 
The enforcement of the PC as additional constraints raises the total energy by the difference 
$\Delta E_c=E_c^{\rm (P)} - E_c^{\rm (E)} $ which will be zero only in those cases where the minimization
without PC enforcement fulfills all the PC. No such case was found, however, in the systems and functionals
we considered. In Table~\ref{tab:corr}, we include the absolute correlation energies $|E_c^{\rm (E)}|$ without the PC, 
as well as $\Delta E_c$ for the different systems and functionals. Enforcing the PC for the Li atom, and for  
He$_2^+$ raises the correlation energy by an amount $\Delta E_{c}$ which
is of the same order of magnitude as the correlation energy $E^{\rm (E)}_{c}$ itself, for almost all functionals. 
For the other two systems, 
the effect of the PC on the correlation energy varies depending on the system and the functional but it is generally 
smaller. In general, however, $\Delta E_{c}$ is quite substantial. As expected, the ML functional which 
was found to violate the PC less than the other functionals yields smaller $\Delta E_{c}$. The trends we observe are similar for the 3-6 and the 3-8 PC. 


\begin{table}
\setlength{\tabcolsep}{8pt}
\begin{tabular}{l|c|c|c|c} \hline\hline
\multicolumn{5}{l}{\em A. For the 3-6 conditions:} \\ \hline
& M\"uller & BBC3 & ML & PNOF0 \\ \hline
Li & 9.6$\cdot10^{-5}$ & 1.7$\cdot10^{-7}$ & 1.1$\cdot10^{-4}$ & 3.6$\cdot10^{-5}$ \\
H$_3$ & 4.2$\cdot10^{-3}$ & 4.8$\cdot10^{-4}$ & 8.5$\cdot10^{-4}$ & 1.1$\cdot10^{-3}$ \\ 
LiH$^+$ & 1.3$\cdot10^{-4}$ & 0 & 3.4$\cdot10^{-5}$ & 1.1$\cdot10^{-6}$ \\
He$_2^+$ & 4.1$\cdot10^{-4}$ & 9.6$\cdot10^{-5}$ & 3.4$\cdot10^{-4}$ & 1.6$\cdot10^{-4}$ \\ \hline
\multicolumn{5}{l}{}\\
\multicolumn{5}{l}{\em B. For the 3-8 conditions:} \\ \hline
 & M\"uller & BBC3 & ML & PNOF0 \\ \hline
Li & 6.1$\cdot10^{-4}$ & 3.0$\cdot10^{-4}$ & 1.5$\cdot10^{-4}$ & 2.2$\cdot10^{-4}$ \\
H$_3$ & 1.1$\cdot10^{-3}$ & 2.9$\cdot10^{-4}$ & 5.0$\cdot10^{-4}$ & 3.6$\cdot10^{-4}$ \\
LiH$^+$ & 9.1$\cdot10^{-4}$ & 4.6$\cdot10^{-4}$ & 7.8$\cdot10^{-4}$ & 9.4$\cdot10^{-4}$ \\
He$_2^+$ & 3.6$\cdot10^{-3}$ & 1.6$\cdot10^{-3}$ & 6.0$\cdot10^{-4}$ & 1.5$\cdot10^{-3}$ \\
\hline\hline
\end{tabular}
\caption{\label{tab:pinned} Deviation from pinning for the largest occupation number, $1-\lambda_1$, with the 
3-6 or 3-8 PC enforced. With the enforcement of the EC alone, this occupation number is pinned to 1 in all
cases considered.}
\end{table}

State pinning, i.e.\ the existence of natural spin-orbitals with border occupations (1 or 0) is very common within
RDMFT approximations. More specifically, with the enforcement of the EC alone there exist many spin-orbitals 
(core states) that are pinned at one while pinning at zero is rather unlikely. i.e.\ there
is an ``over-pinning'' of core states at one for many approximations. However, this inaccuracy is 
not a severe flaw in most cases, since these natural spin-orbitals have an occupation very close to one anyway, 
and they are usually core states not affecting the chemical bonding.
Although the existence of pinned states in the exact 1-RDM of interacting electron systems is not an easy question,
simple 3-electron systems like those studied in this work should not have pinned states. 
However, with only the EC enforced, for all the functionals employed here $\lambda_1=1$ for all systems.
Interestingly, as we see in Table~\ref{tab:pinned}, applying the 3-6 PC in addition, $\lambda_1$
becomes fractional in all cases apart from one (LiH$^{+}$ with the BBC3 functional). Moreover, imposing the 3-8 
conditions leaves no pinned states and in most cases unpins $\lambda_1$ even more than the 3-6 conditions for the 
corresponding system and functional. This is due to enforcing condition 5 which, as discussed above, cannot be satisfied if $\lambda_1=1$ unless $\lambda_2=\lambda_3$. As the latter is not the case in any of our systems the only way to satisfy this condition is the unpinning of $\lambda_1$. 


\begin{table}
\setlength{\tabcolsep}{4pt}
\begin{tabular}{l|cc|cc|cc|cc} \hline\hline
\multicolumn{9}{l}{\em A. For the 3-6 conditions:} \\ \hline
& \multicolumn{2}{c|}{M\"uller}  & \multicolumn{2}{c|}{BBC3 } & 
\multicolumn{2}{c|}{ML } & \multicolumn{2}{c}{PNOF0 } \\
& w/o & with & w/o & with & w/o & with & w/o & with\\ \hline
Li & 13.42 & 0.11 & 11.64 & 0.09 & 0.17 & 0.08 & 0.61 & 0.04 \\
H$_3$ & 10.89 & 9.85 & 6.41 & 5.26 & 0.52 & 0.49 & 1.93 & 1.91 \\ 
LiH$^+$ & 1.04 & 0.33 & 1.09 & 0.15 & 0.18 & 0.12 & 0.42 & 0.17 \\
He$_2^+$ & 14.56 & 3.72 & 13.78 & 2.86 & 0.54 & 0.35 & 2.26 & 1.20 \\
\hline
\multicolumn{9}{l}{}\\
\multicolumn{9}{l}{\em B. For the 3-8 conditions:} \\ \hline
& \multicolumn{2}{c|}{M\"uller } & \multicolumn{2}{c|}{BBC3 } & \multicolumn{2}{c|}{ML } & \multicolumn{2}{c}{PNOF0 } \\
& w/o & with & w/o & with & w/o & with & w/o & with\\ \hline
Li      & 22.23 & 0.29 & 15.53 & 0.12 & 0.35 & 0.18 & 1.36 & 0.10\\
H$_3$   & 11.79 &11.30 &  6.50 & 5.33 & 0.63 & 0.62 & 2.22 & 2.21  \\ 
LiH$^+$ &  1.51 & 0.67 &  1.24 & 0.18 & 0.27 & 0.25 & 0.53 & 0.33\\
He$_2^+$& 15.03 & 4.23 & 13.85 & 2.79 & 0.64 & 0.46 & 2.34 & 1.33  \\
\hline\hline
\end{tabular}
\caption{\label{tab:weak_occ} Sum of the occupation numbers ($\times$100) of weakly occupied orbitals with or without enforcing the 3-6 or 3-8 PC.}
\end{table}

In addition to artificially pinning states, many approximate functionals are also known to overestimate the occupation of the weakly occupied states \cite{L2013}. Due to the unpinning of the largest occupation number, one might expect that the occupation of those states is further increased. However, for both the 3-6 conditions and the 3-8 conditions (see Table~\ref{tab:weak_occ}) this is clearly not the case. For all systems, and all 
functionals, the occupation of the weakly occupied orbitals is actually decreased in some cases even quite significantly. 

\begin{widetext}
\begin{table}
\begin{tabular}{l|c|c|c|c}\hline\hline
         & M\"uller    & BBC3            & ML            & PNOF0\\ \hline
Li       & 1, 5        & 2, 5, 6, 16     & 5             & 1, 5\\ \hline
H$_3$    & 2, 5, 7, 11,& 2, 5, 7, 11,    & 5, 11, 18, 26 & 4, 7, 11, 18, \\
         & 18, 25, 26  & 18, 25, 26      &               & 23, 25 \\ \hline
LiH$^+$  & 1, 2, 4, 5, & 1, 2, 3, 4,     & 2, 6, 10      & 2, 3, 6, 7\\ 
         & 7, 18       & 5, 6, 7, 8,     &               &           \\
         &             & 10, 12, 14, 16, &               &           \\
         &             & 17, 18, 31      &               &           \\ \hline
He$_2^+$ & 2, 5, 7, 18 & 2, 5, 7, 18     & 2, 5, 7, 18   & 2, 5, 7, 18 \\ 
\hline\hline
\end{tabular}
\caption{\label{tab:boundary} Saturated 3-8 conditions, i.e.\ those conditions where the set of the
optimal occupations lies on the boundary of the polytope when enforced. 
For the numbering convention see Table~\ref{tab:38const}.}
\end{table}
\end{widetext}

Since all the 3-8 constraints are inequalities one distinguishes between those constraints that lie on the boundary of the polytope when being enforced, i.e.\ where the lhs.\ is equal to the rhs.\, and those constraints that lie inside the polytope, i.e.\ where the lhs.\ is truly smaller than the rhs. Physically, being on the boundary implies that any further reduction of the total energy will come at the cost of violating at least one of these constraints. In Table \ref{tab:boundary}, we list those conditions that are lying on the boundary for each system and functional. As one can see, the smallest number of boundary constraints appears for the ML functional which also showed the smallest violation of the constraints when they were not enforced.

\section{Conclusions and Outlook}\label{sec:conclusion}
In conclusion, we examined if the additional enforcement of the pure-state $N$-representability conditions, also known 
as generalized Pauli constraints, has an impact on RDMFT calculations. More specifically, 
 we checked if the standard RDMFT optimization for a few representative approximations under the EC of 
Coleman respect the PC conditions for 3-electron systems, namely the Li atom, the linear Hydrogen trimer, and the positive ions of the He$_2$ and LiH molecules. We found that in all cases at least some of the pure-state conditions are violated. We then
applied the 3-6 and 3-8 PC as additional constraints in the minimization and found that the increase in the
total energy corresponds to a significant fraction of the correlation energy obtained with the EC alone. In addition, we found 
that the optimal 1-RDM is qualitatively improved by the  enforcement of 
PC since the pathology of state pinning is alleviated and the total charge occupying weakly occupied orbitals
is reduced. Finally, we found that, in analogy to state pinning of the EC, many PC are saturated, i.e.\ the optimal
1-RDM lies on several boundaries of the polytope of pure-state $N$-representability. Thus, in conclusion, we can claim 
that the enforcement of PC has a significant impact on the results of RDMFT approximations that could potentially be
explored to define better approximations.

Unfortunately, the number of $N$-$M$ PC for large number of electrons $N$ and sizes of the Hilbert space $M$ can be 
quite large and difficult to handle in practical implementations. However, the existence of a systematic way to derive
these conditions for arbitrary $N$ and $M$ and their linear form should facilitate their adoption in more general applications of RDMFT in the future. 

\begin{acknowledgments}
IT and NH acknowledge support from
a Emmy-Noether grant from Deutsche Forschungsgemeinschaft. 
NNL acknowledges support the Greek
Ministry of Education (E$\Sigma\Pi$A program),
GSRT action $\rm KPH\Pi I\Sigma$,
project ``New multifunctional Nanostructured Materials and Devices - POLYNANO'' (No. 447963).
\end{acknowledgments}

\bibliographystyle{apsrev}

\end{document}